\documentclass[seceq]{ptptex}
\usepackage{wrapft}

\usepackage{graphicx}

\title{%        %You can use \\ for explicit line-break.
Coherent photonuclear reactions for isotope transmutation  
}
\author{%       %Use \scshape  for the family name.
Hiroyasu \textsc{Ejiri}$^{1,}$\footnote{ejiri@rcnp.osaka-u.ac.jp} and S. \textsc{Dat\'e}$^2$ 
}

\inst{%         %Affiliation, neglected when [addenda] or [errata].
$^1$  Research Center for Nuclear Physics, Osaka University, Osaka 567-0047, Japan,\\
Nuclear Science, Czech Technical University, Brehova, Prague, Czech Republic,
\\
$^2$ Japan Synchrotron Radiation Research Institute, Sayo, Hyogo 679-5143}

\recdate{%      %Editorial Office will fill in this.
Feb. 21, 2011}

\abst{%         %This abstract is neglected when [addenda] or [errata].
Coherent photonuclear isotope transmutation (CPIT) produces exclusively radioactive isotopes (RIs) by coherent photonuclear ($\gamma $,n) and ($\gamma $,2n) reactions via E1 giant resonances. Photons to be used are medium energy ($E(\gamma ) \approx $ 12-25 MeV)  photons produced by laser photons backscattered off GeV electrons. The cross sections are as large as 0.2 - 0.6 b (10$^{-24}$cm$^2$), being independent of individual nuclides. A large fraction ($\sim 5\%$) of photons is effectively used for the photonuclear reactions, while the scattered GeV electrons remain in the storage ring to be re-used. CPIT with medium energy photons around 10$^{12-15}$/sec provides specific/desired RIs with the rate of 10$^{10-13}$/sec and the RI density around 0.05-50 G Bq/mg for nuclear science, molecular biology and for nuclear medicines. \\

\textit{Key wards}: RI productions, laser photons, Compton backscattering, GeV electrons, photonuclear reactions, EI giant resonance, nuclear medicine, $^{99}$Mo/$^{99m}$Tc. SPECT/PET}  %Input the subject index(es) of your paper, 
\begin{document}
\maketitle

The present letter aims to report that CPIT (coherent photonuclear isotope transmutation) is quite powerful for exclusive RI (radio isotope) productions (transmutations). Nuclear reactions used for CPIT are coherent photonuclear reactions through giant resonances (GR) by means of laser electron photons, i.e. medium energy photons produced by laser photons backscattered off energetic GeV electrons in a storage ring.  
CPIT is shown to be a very efficient and realistic way to provide various kinds of RIs to be used for nuclear physics, molecular biology, nuclear medicine and for other basic and applied science. 

So far, (n,$\gamma $) reactions and nuclear fissions have been extensively used for RI  productions and transmutations. They are caused by the strong (nuclear) interaction, while the photonuclear reactions are by EM interactions. Thus the photonuclear reaction  cross section is in general much smaller than typical nuclear cross sections because of the small EM coupling constant. 

Low energy thermal neutrons used for (n,$\gamma $) reactions and/or nuclear fissions are easily obtained by using intense medium energy protons and/or high flux nuclear reactors. On the other hand medium energy photons required for photonuclear RI productions are hardly obtained by conventional methods. 

RIs produced by (n,$\gamma $) reactions and those by nuclear fissions are limited to those with large neutron capture cross sections and those with large fission branches, respectively. Many kinds of fission product RIs, however, are produced in addition to the specific isotope of interest, and thus chemical separation is indispensable for extracting the desired isotope.

Laser electron photons for RI productions have been discussed, as given in recent reports and references therein \cite{eji10, szp10, hab10}. Photonuclear reactions and photofissions for medical isotope productions and nuclear transmutations were evaluated by using the FLUKA simulation code \cite{szp10}. Photonuclear reactions with photon beams of large brilliance and small band width ($\Delta E/E$ = 10$^{-3}$) were discussed for production of medical RIs with high specific activity \cite{hab10}. Isotope productions by using bremsstrahlung photons from medium energy electron beams were discussed in the recent report \cite{bun10}. 

In what follows, we describe briefly unique features of CPIT with laser electron photons to provide efficiently various kinds of RIs, including  $^{99}$Mo/$^{99m}$Tc isotopes and other SPECT/PET RIs, with large RI production rate and high RI density for RIs of interest and little extra RIs.\\ 

%%%%%%%%%%%%%%%%%%%%%%%%%%%%%%%%%%%%%%%%%%%%%%%%%%55
%%%%%%%%%%%%%%%%%%%%%%%%%%%%%%%%%%%%%%%%%%%%%%%%%%

Coherent photonuclear reactions via the isovector E1 giant resonance are used for CPIT. Merits of the coherent reactions are as given below.

1. The cross section is quite large because of the coherent excitation of many nucleons involved in the GR. The energy integrated cross section is given by 

\begin{equation}
\int \sigma (E(\gamma)) dE(\gamma) = 2\pi ^2~\alpha ~\frac {\hbar ^2}{M} ~\frac {N Z}{A}(1+\kappa ) = 270 ~\alpha ~A ~\rm{fm^2~MeV},   
\end{equation}
where $N$ and $Z$ are the proton and neutron numbers, $A=N+Z$ is the mass number, $\kappa \approx 0.3$ is the correction coefficient for the exchange current, and $\alpha = e^2/\hbar \approx$1/137 is the EM coupling constant. 

\begin{table}[h]

\caption{Cross sections for photonuclear reactions at E1 GR and cross sections for electron positron pair creations and Compton scatterings in unit of b=10$^{-24}cm^2$.
}
\vspace{0.3cm}
\begin{center}
\let\tabularsize\normalsize
\begin{tabular}{cccc} 

\hline
Isotope	& $\sigma $(GR) b	& $\sigma $(e) b &	$\sigma (GR)/\sigma (e)$\\
\hline
$^{27}$Al	& 0.08 &0.95 &	0.09\\
$^{63}$Cu	& 0.19	&3.5 &	0.05\\
$^{124}$Sn	& 0.37	& 8.5 &	0.04\\
$^{208}$Pb	& 0.62	& 19.5	& 0.03\\

\hline

\end{tabular}

\end{center}           
 \end{table} 
 
Using the Breit-Wigner resonance shape with the observed width of $\Gamma \approx $ 4.5 MeV for the E1 GR, the cross section at $E(\gamma)=E(GR)$ is expressed as 

\begin{equation}
\sigma(GR) \approx  3 \times 10^{-3} ~A~\rm{b}. 
\end{equation}
                          
The cross section amounts to about 30 $\%$ of the geometrical nuclear cross section. GR resonance cross sections for typical nuclei are shown in Table 1. 
The cross section is proportional to the mass number $A$ because of the coherent excitation. Therefore the small EM coupling of $\alpha \approx $ 1/137 is well compensated by the large factor of $A= 60 \sim $200 in case of medium heavy nuclei.

\begin{figure}[h]
\begin{center}
\includegraphics[width=0.5\textwidth]{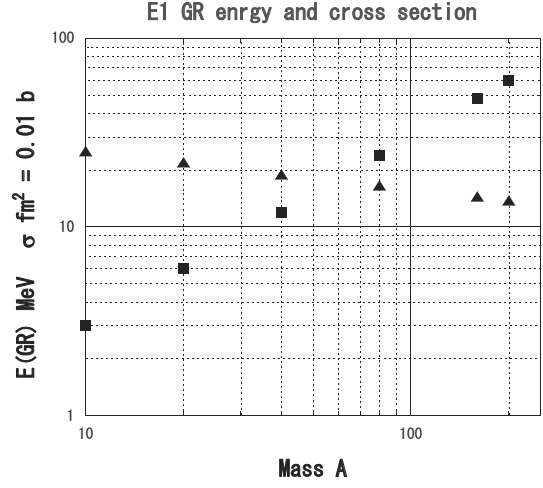}
%\includegraphics[width=0.5\textwidth]{EjiriFig1.pdf}
\caption{ E1 giant resonance (GR) energy (triangles) and photonuclear cross sections via E1 GR (squares) as a function of the mass number $A$
}
\label{fig:1}
\end{center}       % Give a unique label
\end{figure}

2. GR is a macroscopic oscillation of a bulk of protons and that of neutrons. Accordingly the cross section per nucleon, the resonance energy and the resonance width do not depend much on individual nuclides and insensitive to the individual nuclear structures. The resonance energy is expressed as $E(GR) \approx  a A^{-1/5} = 22 \sim 14$ MeV for $A = 30 \sim 200$ nuclei, and the width is as broad as $\Gamma $ = 4 $\sim $5 MeV, as shown in Fig.1. Thus one can preferentially excite GR in any nuclei by using medium energy photons.  This is in contrast to (n,$\gamma $) reactions by thermal neutrons, where the cross section depends much on individual nuclides and nuclear level structures. 

3. Photonuclear reactions on medium heavy nuclei at the GR region are mainly  ($\gamma $,n) and ($\gamma $,2n) reactions, depending on $E(\gamma ) \geq B_n$ or $E(\gamma) \geq B_{2n}$, where $B_n$ and $B_{2n} \approx 2B_n$ are the one and two neutron binding energies, respectively. Then one gets exclusively nuclei with ($Z,N-1$) and ($Z,N-2$) from target nuclei with ($Z,N$) by using the ($\gamma $,n) and ($\gamma $,2n) reactions with adequate energy photons, respectively. In case of light nuclei, $(\gamma $,p) reactions are also used to produce RIs with ($Z-1,N$). Accordingly, it is possible to get specific/desired RI isotopes from stable isotopes, or short-lived RI isotopes from long-lived RI isotopes without producing many extra RIs. This is in contrast to fission products in nuclear reactors, where many other isotopes are necessarily produced.\\

%%%%%%%%%%%%%%%%%%%%%%%%%%%%%%%%%%%%%%%%%%%%%%%%%%%%%%%%%%%%%%%%%%%%%%%%%
%\subsection{ Laser electron photons for photonuclear isotope transmutation}
  
Medium energy photons with the appropriate $E(\gamma) \approx E(GR)$ and $\Delta E \approx \Gamma $ are obtained by Compton backscattering of laser photons from GeV electrons in a storage ring. Laser photons are amplified in energy via scattering off GeV electrons by many orders of magnitude, depending on the electron energy and the scattering angle.
The scattering process of laser photons is schematically shown in Figure 1. 

\begin{figure}[h]
\begin{center}
\includegraphics[width=0.5\textwidth]{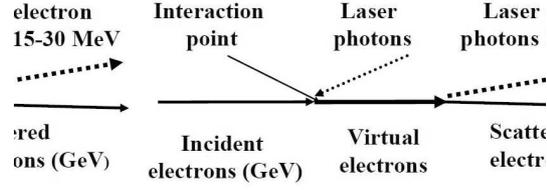}
%\includegraphics[width=0.5\textwidth]{EjiriFig2.pdf}
\caption{Schematic view of production of laser electron photons by Compton backscattering of laser photons from GeV electrons.}.

\label{fig:2}
\end{center}       % Give a unique label
\end{figure}

Here the GeV electron is excited to a virtual state by the laser photon capture, and is split into the scattered GeV electron and the laser electron photon, i.e. the scattered photon. The scattered photon energy $E(\gamma)$ is expressed in terms of the laser photon energy $E(l)$, the scattering angle $\theta $ with respect to the electron beam, and the electron mass $m$ as

\begin{equation}
E(\gamma) = \frac {4~\gamma _e^2~E(l)}{1+4~\gamma _e~E(l)/m~+~\gamma _e^2~\theta ^2}, 
\end{equation}
where $m$ denotes the electron mass and $\gamma _e = E(e)/m$ is the 
Lorentz factor of the incident electron with the energy $E(e)$. Noting $4\gamma _eE(l)/m \ll 1$ in the present case, the photon energy can be expressed as
\begin{equation} 
E(\gamma) = k~E(l)~f(\theta ),                           
\end{equation}
where $k \approx 4 \gamma _e^2$ is the amplification factor, and 
$f(\theta)~\approx ~[1+~\gamma _e^2~\theta ^2]^{-1}$ stands for the angle dependence. 
Since $f(\theta )$ = 1 at the forward angle of $\theta $= 0 , i.e. the backward angle with respect to the incident laser photon, the laser electron photon has the maximum energy of $k~E(l)$ with the amplification factor $k \approx 4 \gamma _e^2$ at $\theta $= 0. 

The energy window of $E(\gamma )-E'(\gamma )$ can be set by using a collimator of an angle aperture of $\pm \Delta \theta $ at the forward direction, where $E(\gamma )$ and $E'(\gamma )$ are the photon energies at the scattering angles of $\theta = 0$ and $\theta = \Delta \theta $.

The laser electron energy and the band width are tuned by adjusting $E(l)$, $E(e)$ and $\Delta \theta  $ so that the energy spectrum  shows the peak at the maximum energy of $E(\gamma ) \approx E(GR) (\sim $15 MeV) and the band width around $\Delta E(\gamma)  \approx \Gamma (\sim $5 MeV).  

The energy amplification factor is as large as $k = 10^{7\sim 8}$ for the GeV electron energy of $E(e) = 1\sim  $3 GeV. Then one gets medium energy photons of $E(\gamma ) \approx $15 MeV to be used for photonuclear reactions by using 1 $\sim $ 0.1 eV laser photons. 

The photon intensity of around 10$^{8-12}$/sec is realistic by using intense lasers and intense electrons. In future high intensity photon sources with the intensity of the order of 10$^{14-15}$/sec will be possible.  \\

The laser electron photons have several advantageous points to charged particles and neutrons. They are as follows.

1. The laser photons are well collimated into a small angular region of $\theta \approx 1/\gamma _e$. The angle spread is around $\theta \approx $ 0.15-0.5 mrad. for 3-1 GeV electrons Thus the nuclear transmutation is confined in the small cylinder within 1 mm in radius at around a few meters from the laser electron interaction point. This makes it possible to get isotopes with very high RI density of the order of G Bq/mg, which is crucial for nuclear physics, molecular biology and nuclear medicines. 
 
2. The large fraction ($\sim 5\%$) of the incident photons are effectively used for the isotope transmutation, as given in Table 1. Here a very thick target of the order of 20$\sim $30 gr/cm$^2$ can be used for photon beams since photons have no electric charge  to interact with atomic electrons. Note that charged particles lose their energy mostly via interaction with atomic electrons without being used for nuclear reactions. 

3. The scattered GeV electrons lose the energy via the interaction with the laser photon only by 0.3 $\sim $1 $\%$ , depending on the scattering angle, and thus most of them can still remain in the storage ring to be re-accelerated up to the original energy by RF power supply. 

4.  The laser electron photons tuned to the GR energy are well above the ($\gamma $,n) threshold energy. Then about a half of the laser electron photons can be used for photonuclear reactions. This is in contrast to bremsstrahlung photons, where the photon yields drop down rapidly with increase of the energy, and thus a small part of the high energy tail can be used for photonuclear reactions \cite{bun10}.
 
5. Some laser electron photons interact with target nuclei to excite GR, which decays by emitting 1 or 2 fast neutrons and a few $\gamma $ rays, while others interact with the electric field of the target nuclei to produce electron positron pairs with $E(e^{\pm})= 5\sim 8$ MeV. These neutrons, $\gamma $ rays and  electron-positron pairs may be used for basic and applied science. Slowing  down the fast neutrons, one may re-use them for another isotope transmutation \cite{szp10}.\\

%%%%%%%%%%%%%%%%%%%%%%%%%%%%%%%%%%%%%%%%%%%%%%%%%%%%%%%%%%
%\subsection{Photonuclear RI productions}

Let's evaluate the RI production rate and the RI density for CPIT. Photonuclear ($\gamma $,n) and ($\gamma $,2n) cross sections via the E1 GR are given as 
$\sigma ^e (n) = \sigma ^e (GR) B(n)$, and $\sigma ^e(2n) = \sigma ^e(GR) B(2n)$,   
where $B(n)$ and $B(2n)$ are the n and 2n branching ratios and $\sigma ^e(GR)$ is the effective cross section in the GR energy window. Practically, $B(n)$ and $B(2n)$ are nearly 1 by selecting the laser electron photon energies as $B(n) \leq E(\gamma) \leq B(2n)$ and 
$B(2n) \leq E(\gamma) \leq B(3n)$. Experimentally, we use photons with the spectrum window of $\Delta E \approx \Gamma $. Thus the effective cross section is $\sigma ^e (xn) \approx 0.8~ \sigma (xn)~=0.8 ~B(xn)~\sigma (GR)$. 

The RI production rate $N(Z,N-x)$ by using the $(\gamma,xn)$ reaction on the $A(Z,N)$ target nucleus is given as 
\begin{equation}
N(Z,N-x) = N'(\gamma)~a~ \sigma ^e(xn) N(T), 
\end{equation}
where $N'(\gamma)$ is the number of photons per sec in the energy window of $\Delta E \approx 5$ MeV, $a$ is the attenuation factor for the incident photons through the thick target, and $N(T)$ is the number of target nuclei per unit area. We evaluate the RI production rates in two cases. 

In case of FEL, the production rate can be evaluated on the basis of the HIGS data of $N(\gamma $) = 2~10$^9$/sec for 50 mA electrons with $E(e)$ = 0.474 GeV and $E(l)$ = 1.6 eV ($\lambda $ =780 nm) \cite{wel09}. 

Extrapolating these data, one may assume/expect to get the total number of photons around 
$N(\gamma) \approx 2 \times 10^{12}$/sec / 500 mA for $E(e)$=1.2 GeV and 
the $N'(\gamma) \approx 10^{12}$ /sec for the photons in the energy window. Using a target with thickness of around 30 gr/cm$^2$ and $a \approx $ 0.62, the RI production rate for a typical case of $A=100,~ B(xn)$ = 0.7 and $\sigma (GR)$ = 0.3 b is given as,

\begin{equation}
N(Z,N-x) \approx 2 \times 10^{10} /sec. 
\end{equation}
                          
The photons are well defined into the small angular region of $\theta = 1/\gamma _e  \approx 4\times 10^{-4}$ radian, i.e. within a radius $r \approx 0.8$ mm at 2 m from the interaction point. Then the target cylinder with 1.6 mm in diameter is 0.6 gr. Thus the RI density after sufficiently long irradiation (multiple of the product half-life) is 2 $\times 10^{10}$ /sec/0.6 gr = 0.033 G Bq/mg, which is equivalent to 0.033 G Bq/ml in case of liquid with 1 gr/cc. Using target nuclei enriched into particular isotopes, one gets efficiently the isotope transmutation, being almost free from other extra/impurity isotopes.

In case of CO$_2$ laser photons injected anti-parallel to a beam of GeV electrons in a storage ring, the number of the backscattered photons per sec is given as \cite{dat10}.
\begin{equation}                 
N(\gamma) = 1\times 10^8~ \frac{I~l~P}{s},  
\end{equation}
where $I$ is the current of the electron beam  and $P$  stands for the power of the laser, respectively in units of A and W. A segment with the length $l$ of the electron beam is assumed to be contained inside the laser beam of the cross section area $s$, respectively in units of m and mm. A finite Rayleigh length makes $s$ dependent on $l$. In case of a Gaussian laser beam, the dependence is written as

\begin{equation}
s = \frac{\lambda }{2} \frac{l}{L}, ~~~~~~~ L~=~arctan (\frac{l \lambda}{2\pi w^2}),                      
\end{equation}
where $\lambda ~(10.6 \mu m$) in unit of $\mu $m and $w$ in unit of mm are respectively the wavelength and the waist radius of the laser beam at the middle of the segment $l$. To enhance $N(\gamma $) , one may increase $l$ and reduce $w$. It is however bounded by a value $N_0(\gamma) = 4.4\times 10^7 ~I~P$, which corresponds to a sufficiently large segment length $l$ compared with the Rayleigh length. When a relationship
$l \geq 4w^2$ is fulfilled, $N(\gamma)$ counts more than 90 $\%$ of $N_0$.

 Low emittance electron beams in the third generation synchrotron light sources allow us to assume $w$ = 0.5 mm for which an adequate segment length is $l$=1 meter. To obtain $N(\gamma) = 10^{12}$/sec with $I$ = 500 mA as considered in the FEL case, the power of the laser is required to be $P$ = 46 kW. To be conservative, we may say that a well established technique of the laser backward scattering utilizing a few kW CO$_2$ laser photons can provide a mean to study nuclear transmutation and RI production with a production rate of an order of $10^{9}$/s.

\begin{table}[h]

\caption{Radioactive isotopes to be used for PET tracers and photonuclear reactions.}
\vspace{0.3cm}
\begin{center}
\let\tabularsize\normalsize
\begin{tabular}{ccccc} 

\hline
Isotope	& Halflife &	$Q(EC)$ keV	& Reaction	& Abundance \\
\hline
$^{11}$C &	20.4 m &	1982 &	$^{12}$C ($\gamma $,n) &	98.9\\
$^{13}$N	& 10.0 m &	2220	& $^{14}$N ($\gamma $,n) &	99.6\\
$^{15}$O	& 2.04 m &	2754	& $^{16}$O ($\gamma $,n) &	99.8\\
$^{18}$F	& 109.8 m	& 1656 &	$^{19}$F ($\gamma $,n) &	100\\
$^{62}$Zn &	9.2 h	& 1627 &	$^{64}$Zn ($\gamma $,2n)	& 48.6\\
$^{68}$Ga &	67.6 m	& 2921	& $^{69}$Ga($\gamma $,n) &	60.1\\
\hline

%%%%%%%%%%%%%%%%%%%%%%%%%%%%%%%%%%%%%%%%%%%%%%%%%%

\end{tabular}

\end{center}           
 \end{table}

\begin{table}[h]

\caption{Radioactive isotopes to be used for SPECT tracers and photonuclear reactions.}
\vspace{0.3cm}
\begin{center}
\let\tabularsize\normalsize
\begin{tabular}{ccccc} 

\hline
Isotope	& Halflife d &	$E$ keV	& Reaction	& Abundance \\
\hline          
            
$^{67}$Ga	& 3.26	& 93, 185 &	$^{69}$Ga ($\gamma $,2n) &	60.1\\
$^{99}$Mo, $^{99m}$Tc &	2.75, 0.25 & 140 &	$^{100}$Mo ($\gamma $,n)	& 9.6\\
$^{111}$In	& 2.8	& 245, 171	& $^{113}$In ($\gamma $,2n)	& 4.3\\
$^{126}$I	& 13	& 389, 666 & 	$^{127}$I ($\gamma $,n) &	100\\
$^{195m}$Pt	& 4.02	& 259	& $^{196}$Pt($\gamma $,n) &	25.4\\
$^{201}$Tl	& 3.04	& 167 &	$^{203}$Tl ($\gamma $,2n) &	29.5\\

\hline

%%%%%%%%%%%%%%%%%%%%%%%%%%%%%%%%%%%%%%%%%%%%%%%%%%

\end{tabular}

\end{center}           
 \end{table}

Intense sources of medium energy photons with $N(\gamma) = 10^{13-15}$/s may provide RIs around $N \approx 10^{11-13}$/s. The RI density is evaluated for a typical case of $E(e)$ = 2 GeV with $\gamma $=3.9 10$^3$. Laser electron photons with $E(\gamma )$=15 - 10 MeV and 24 - 16 MeV for $(\gamma $, n) and ($\gamma $,2n) reactions are obtained, respectively, by scattering of 0.24 eV and 0.39 eV laser photons at $\theta = 0$ - 0.16 m rad. Thus the colimater to be used for both reactions are the one with apperture of $\Delta \theta = \pm $ 0.16 m rad. i.e. the radius of 0.65 mm at 4 m from the interaction point. Using a 30 gr/cm$^2$ target,    
$N\approx 10^{11-13}$/s  can be confined in a sylinder of $r$ =0.65 mm with the density of 0.5 - 50 G Bq/mg after long irradiation.

Let us discuss the $^{99}$Mo/$^{99m}$Tc isotopes, which are widely used as SPECT isotopes. The $^{99}$M isotopes are produced by $^{100}$Mo($\gamma ,$n) reactions and also $^{100}$Mo($\gamma ,$p) reactions followed by $\beta ^-$ decays. CPIT with $N(\gamma) \approx $ 10$^{13-15}$/s and enriched $^{100}$Mo isotopes provides the RIs of $^{99}$Mo and $^{99m}$Tc with the production rate of 10$^{11-13}$/sec and the RI density of 5 10$^{8-10}$ G Bq/mg after sufficiently long irradiation. \\

%%%%%%%%%%%%%%%%%%%%%%
%%%%%%%%%%%%%%%%%%%%%%%%%%%%%%%%%%%%%
%\section{Concluding remarks and perspectives on photonuclear reactions}

CPIT with large efficiency is quite attractive from ecological view points.  
Since the GeV electrons stored in a storage ring lose little their energy via interactions with laser photons, they remain in the ring. The laser electron photons are efficiently used for production of desired RIs. Then the overall efficiency of the RI production is many orders of magnitude larger than that of the charged particle accelerators and nuclear reactors. Fast neutrons are used for nuclear transmutation \cite{nag08}, but only a small fraction of the charged particles are used to produce the fast neutrons. Intense electron accelerators provide bremestrahlung photons for photonuclear isotope transmutation \cite{bun10}, but most of photons are below the threshold energy of the photonuclear reaction. 
 
There are several programs of intense photon sources under progress \cite{szp10, hab10, bar10}. MAGa-ray is the project at LLNL for high intensity photons with 10$^{12}$ photons/sec in the MeV region and ELI-NP is the one at Romania for higher energy photons with 10$^{13}$ photons/sec in the GR energy region of $E(\gamma) \geq $ 19.5 MeV.  They plan to achieve the intensity of around 10$^{15}$/sec.  In Tokai Japan, the ERL(Energy Recovery Liniac) project is under progress to provide intense photons for resonance fluorescence \cite{haj09}. 

If the natural abundance of the target nuclide is not large, enrichment in the specific isotope is effective to increase the efficiency by the enrichment factor and to reduce other RI productions. Tone-scale enrichment into $^{100}$Mo for $^{99}$Mo/$^{99m}$Tc and other isotopes are realistic by means of centrifugal isotope separators. Such isotopes are also used for basic science such as neutrino studies by double beta decays \cite{eji05}. A large scale isotope separation plant with the separation rate of the order of 100 kg/year is interesting for basic and applied science.

It is noted that CPIT can be used to detect small impurities of specific stable and radioactive isotopes. They are excited by photonuclear reactions via E1-GR, and are detected by measuring prompt and delayed gamma rays characteristic of the isotopes of the order of ppb/ppt levels. Thus it can be applied for nondestructive impurity and radionuclide assay.   

In short, CPIT with laser electron photons provides exclusively various kinds of specific/desired isotopes with the large production rate and the high density for basic and applied science. CPIT is of potential interest also for nuclear transmutation of long-lived nuclei.  Accordingly, CPIT using intense laser photons and GeV electrons in a storage ring is quite complementary to other methods using high flux reactors and intense charged particle accelerators. 

It is of great interest from scientific and ecological view points to build high intensity electron storage rings combined FEL to provide medium energy photons with intensity of the order of 10$^{13-15}$/sec. Such intense photons can be widely used for producing byproduct fast neutrons, $\gamma $ rays and electron positron pairs as well as for the photonuclear RI productions for basic and applied science.

\section*{Acknowledgement}
We thank Prof. Y. Asano, , Prof. Date, Prof. C. Rangacharyulu and Prof. B. Szpunar for valuable discussions.

\end{document}